\documentstyle[11pt,aaspp4]{article}
\received{24 April 1996}
\input tex.def
\input psfig.tex

\slugcomment{To Appear in {\it The Astrophysical Journal}}

\begin{document}

\title{High-Dispersion Spectroscopy of Luminous, Young Star Clusters: 
Evidence for Present-Day Formation of Globular Clusters
\footnote{Based on observations obtained at the W. M. Keck Observatory.}}

\author{Luis C. Ho}
\affil{Harvard-Smithsonian Center for Astrophysics, Cambridge, MA 02138}

\and

\author{Alexei V. Filippenko}
\affil{Department of Astronomy, University of California, Berkeley, CA 94720-3411}

\begin{abstract}
We present evidence that some of the compact, luminous, young star 
clusters recently discovered through images taken with the {\it Hubble 
Space Telescope (HST)} have masses comparable to those of old Galactic globular 
clusters.  The ``super star cluster'' in the center of the nearby amorphous
galaxy NGC~1705 has been observed with high dispersion at optical wavelengths 
using the HIRES echelle spectrograph on the Keck 10~m telescope.  Numerous 
weak metal lines arising from the atmospheres of cool supergiants have 
been detected in the integrated spectrum, permitting a direct measurement of 
the line-of-sight stellar velocity  dispersion through cross-correlation 
with a template star; the result is $\sigma_*$ = 11.4$\pm$1.5 \kms.
Assuming that the system is gravitationally bound and using a cluster size 
measured from {\it HST} images, we apply the virial theorem to obtain the 
dynamical mass.  Its derived mass [(8.2$\pm$2.1)\e{4} \solmass], mass density 
(2.7\e{4} \solmass\ pc$^{-3}$), and predicted mass-to-light ratio after aging 
by 10--15 Gyr [0.6--1.6 ($M/L_V$)$_{\odot}$] closely resemble those of 
the majority of evolved Galactic globular clusters.  The central cluster in 
NGC~1705 appears to be very similar in nature to one of the bright clusters 
in NGC~1569, which was discussed earlier this year by Ho \& Filippenko.  We 
also observed 
the brightest cluster in the Magellanic irregular galaxy NGC~4214, but no lines 
suitable for measuring its velocity dispersion were detected, most likely 
because of the very young age of the cluster. Although these observations need 
to be extended to a much larger sample of objects before generalizations can 
be made concerning the nature of similar clusters observed in other galaxies, 
our preliminary results are tantalizing and strongly suggest that, at least in 
two cases, we are witnessing globular clusters in their extreme youth.
\end{abstract}


\section{Introduction}

Globular clusters are believed to be among the oldest relics in galaxies;
as such, understanding the circumstances and physical mechanisms by which 
they formed may shed light on the process of galaxy formation itself (see 
contributions in Smith \& Brodie 1993).  A number of lines of evidence suggest 
that the stars within most globular clusters must have formed coevally over 
a short duration (Murray \& Lin 1993).  Whatever the exact conditions might 
have been, by all accounts such systems represent a highly extreme mode of 
star formation that is evidently unseen at the present epoch (Harris 1991), 
with the possible exception of a few of the young and intermediate-age 
populous clusters in the Magellanic Clouds (Mateo 1993).  

Thus, there has been great 
excitement surrounding the discovery, from images taken with the {\it Hubble 
Space Telescope (HST)}, of a class of luminous, compact, and 
apparently young star clusters in the cooling-flow/merger-remnant galaxy 
NGC~1275 (Holtzman \etal 1992), which was shortly thereafter followed by other 
similar examples found in the merging galaxies NGC~7252 (Whitmore \etal 1993) 
and NGC~4038/NGC~4039 (Whitmore \& Schweizer 1995).  The very high 
luminosities and small sizes of these objects, coupled with their relatively 
young ages, prompted suggestions that they may be young globular clusters.  
Numerous examples of such young clusters have surfaced from other {\it HST} 
imaging studies, both at optical and ultraviolet (UV) wavelengths (see Ho 1996 
for a summary).  Although some of the most spectacular cases have been 
found in interacting and merging systems, it now appears that these clusters 
can also form in more quiescent environments such as circumnuclear rings in 
relatively undisturbed galaxies (Benedict \etal 1993; Barth \etal 1995, 1996; 
Maoz \etal 1996).  Evidently the central regions of certain dwarf irregular 
and amorphous galaxies are also conducive to forming compact, luminous 
clusters.  Some, like NGC~1569 and NGC~1705 (O'Connell, Gallagher, \& Hunter 
1994), appear isolated, while others such as NGC~1140 (Hunter, O'Connell, \& 
Gallagher 1994), M82 (O'Connell \etal 1995), and NGC~4214 (Leitherer \etal 
1996) show obvious signs of ongoing interaction.  Several seminal ground-based 
studies (O'Connell \& Mangano 1978; Arp \& Sandage 1985; Melnick, Moles, \& 
Terlevich 1985; Lutz 1991) have adumbrated these findings, but it took the 
resolving power of {\it HST} to showcase the extraordinary properties of the 
clusters.  

Despite the recent progress, many basic properties of the clusters 
remain unknown.  At the moment, perhaps the most pertinent missing piece of 
information to test the young globular cluster hypothesis is the 
mass of the clusters.  Although the masses of globular clusters in the 
Galaxy span a wide range, the average mass ($\sim$2\e{5} \solmass; Mandushev, 
Spassova, \& Staneva 1991) greatly exceeds that of open clusters.  
For instance, even a fairly rich open cluster such as M11 has a mass of only 
$\sim$5000 \solmass\ (Mathieu 1984).  The mass of the central portion of the 
M35 cluster has been estimated to be $\sim$2000--3000 \solmass\ 
(Leonard \& Merritt 1989), while the Hyades contain no more than $\sim$400 
\solmass\ (Gunn \etal 1988).  Thus, in general there should be no 
ambiguity between the masses of open and globular clusters.  A more robust 
discriminant might be the mass density, since globular clusters tend to be 
much more compact than open clusters.  A reliable estimate of both the size 
and mass of the young clusters should therefore clarify 
the nature of these objects.  Although the accuracy of the published cluster 
sizes can certainly be improved, fairly stringent measurements or upper limits 
are already available from {\it HST}.  By contrast, no {\it direct}, 
model-independent mass estimates exist.  Masses quoted in the literature 
invariably rely on stellar population models, whose uniqueness and reliability 
are difficult to judge given their dependence on a large number of poorly 
constrained parameters.

From a practical viewpoint, the clusters found in the nearby dwarf
galaxies offer us the best chance to study the phenomenon in detail.  In 
particular, NGC~1569 hosts two dominant clusters (conventionally 
denoted ``A'' and ``B''; see Arp \& Sandage 1985), while NGC~1705 contains 
only one (Melnick \etal 1985a).  Because of their proximity, these three 
clusters have been spatially 
resolved to a greater degree than any others studied with {\it HST}
(O'Connell \etal 1994).  The size of the central knot in NGC~4214 is somewhat 
less certain (Leitherer \etal 1996), but nevertheless better known than those 
in more distant galaxies.  We take advantage of the well-determined sizes and 
the relative brightnesses and isolation of these clusters to try 
to obtain velocity dispersions and estimate their dynamical masses.  

Using the Keck 10~m telescope on Mauna Kea, Hawaii, we acquired 
high-dispersion optical spectra of one of the two clusters in NGC~1569 
(object ``A,'' hereafter NGC~1569-A), and of the single dominant cluster 
in NGC~1705 (hereafter NGC~1705-1) and in NGC~4214 (hereafter NGC~4214-1).  
This paper will primarily concentrate on NGC~1705-1 and NGC~4214-1; Ho \& 
Filippenko (1996; hereafter Paper I) discuss the results on NGC~1569-A.  Some 
observational and technical details not given in Paper I are described here.

\section{Observations and Data Reduction}

Object selection was guided primarily by sensitivity 
considerations and by the ability to spatially isolate the cluster using a 
narrow slit.  In practice, we were thus limited to the most nearby systems.  
In order to facilitate background subtraction, preference was given to
clusters with relatively uncomplicated surroundings, since the spectrograph
used (see below) did not yet have an image rotator.  Good data were taken 
for three clusters having $V\,\approx$ 14.7--14.8 mag (see \S\ 1 and Table 1); 
regrettably, time limitations did not permit observation of cluster B in 
NGC~1569, whose photometric properties appear very similar to those of 
NGC~1569-A (O'Connell \etal 1994).  We also attempted to obtain spectra of the 
brightest cluster in NGC~1275 
($V$ = 18.2 mag; Holtzman \etal 1992) for which Zepf \etal (1995) had recently 
acquired a low-resolution spectrum.  The bright galaxy background in this 
case greatly complicated centering and guiding of the cluster on the slit, and 
our data turned out to be of poor quality; we will not further consider 
this object here.

The spectra were acquired on 9 January 1996 UT using the HIRES echelle
spectrograph (Vogt \etal 1994) on the Keck 10~m telescope.  The echelle and 
cross-disperser grating angles were carefully tuned to select the spectral 
regions of interest, since the CCD chip does not cover the full spectral 
range for 
\lamb\ \gax 5100 \AA.  Two setups were designed for this experiment, both 
using the red collimator of HIRES.  The ``optical'' setting extended from 
$\sim$3900 \AA\ (to include Ca~II H \& K) to $\sim$6280 \AA\ in 34 spectral 
orders; it was selected to sample as many metallic and molecular 
stellar absorption features as possible in the violet to red region.  A KV380 
order-blocking filter was used to cut off second-order blue light.  The 
``near-infrared'' (NIR) setting, on the other hand, was primarily intended to 
record the Ca~II IR triplet (\lamb\lamb 8498, 8542, 8662); the wavelength 
coverage extended from $\sim$6500 \AA\ to 8800 \AA\ in 16 orders, with 
significant gaps, and an OG530 order-blocking filter was used.
NGC~1705-1 and NGC~4214-1 were observed in both settings, whereas NGC~1569-A 
was observed only in the optical setting.

The length of the slit for the optical setting was 7\asec\ (``C5'' decker; 
Vogt 1994), and 14\asec\ for the NIR setting (``D1'' decker).  We chose these 
values to simultaneously maximize the amount of sky coverage while preventing 
significant overlap among the orders; in the case of the optical setting, 
however, the first 5 or 6 orders overlapped partially.  The bulk of the 
observations were taken through a slit of width 1\farcs15, yielding an 
effective spectral resolution, as determined from the profiles of the 
comparison 
lamp emission lines, of $R\,\approx$ 38,000 ($\sigma$ = 3.4 \kms).  If the 
clusters indeed have masses comparable to those of most globular clusters 
(5\e{4}--5\e{5} \solmass; Mandushev \etal 1991), we expect velocity 
dispersions along the line of sight to be $\sigma_*\,\approx$ 5--15 \kms\ 
(e.g., Illingworth 1976; Mandushev \etal 1991; Dubath, Mayor, \& Meylan 1993), 
which will be detected comfortably with this resolution.  On the other hand, 
we do not know {\it a priori} the masses of the clusters; to ensure 
sensitivity to masses as small as, say, 10$^4$ \solmass, the typical 
cluster radius of $\sim$2 pc implies that a velocity resolution of $\sigma_*\, 
\approx$ 3 \kms\ must be achieved.  Thus, for NGC~1705-1, we also obtained 
limited data using a slit of width 0\farcs57 (deckers ``B2'' and ``B3'' 
for the optical and NIR settings, respectively), with an effective resolution 
of $R\,\approx$ 60,000 ($\sigma$ = 2.1 \kms).  Note that the chosen slits 
(especially the wide one) sample most of the light from the clusters in the 
absence of atmospheric seeing, provided that the objects are properly centered 
and the telescope guiding is stable, since the half-light diameters of 
2--4 pc subtend only 0\farcs1--0\farcs3 on the sky.

We binned the Tektronix chip (2048 $\times$ 2048 square pixels) 1 $\times$ 2 
(spectral $\times$ spatial dimension) in order to decrease the effective 
readout noise; each binned pixel projected to 0\farcs41 on the sky, with 
a readout noise of $\sim$5 electrons.  Bias frames taken at the beginning and 
end of the night indicate that the chip is stable and has negligible 
dark current.

NGC~1569-A stood out prominently against the galaxy background in the 
acquisition/guiding camera; cluster B, of slightly lower brightness, is 
located approximately 8\asec\ away (O'Connell \etal 1994) and did not present 
a source of confusion.  We integrated on the cluster for a total of 7200 s in 
the optical setting, split into 4 half-hour exposures to facilitate the 
removal of cosmic rays.  Being by far the brightest object at visual 
wavelengths in NGC~1705, NGC~1705-1 was also readily identified and centered 
on the slit.  According to Meurer \etal (1995), a second cluster (NGC~1705-2)
is located 0\farcs94 away; however, since it is fainter than the primary object 
by 3 magnitudes (in the UV), it is unlikely to contaminate significantly the 
signal of NGC~1705-1.  The most useful data resulted from a total integration 
of 3600 s (split into 2 half-hour exposures) in the optical setting with the 
wide slit.  We also took spectra in the higher resolution mode at both the 
optical (2 $\times$ 300 s) and NIR (2 $\times$ 600 s) settings, but the 
integration times were insufficient to yield useful results.  Finally, 
NGC~4214-1 was observed for 5200 s (3 half-hour exposures) and 2400 s 
(2 1200-s exposures) in the optical and NIR settings, respectively.  Line 
emission 
from H~II regions contributes substantially to the optical flux of NGC~4214, 
especially in the immediate vicinity of the star cluster of interest 
(Sargent \& Filippenko 1991).  To minimize confusion in the identification of 
the object and subsequent guiding, we found it useful to introduce a (largely) 
line-free color filter in the light path of the acquisition/guiding camera.

To enable identification and removal of telluric lines in the spectra of the 
program objects, we observed the sdF5 stars HD~19445 and HD~84937 (Oke 
\& Gunn 1983), whose spectra have relatively few features.  In addition, we 
took short exposures of several bright template stars with known spectral 
types, selected from the Bright Star Catalog, for use in the cross-correlation 
method to derive velocity dispersions.  As we are
primarily interested in kinematic information, no effort was made to perform
flux calibration.  The program objects and standard stars were interleaved with 
exposures of thorium and argon hollow cathode comparison lamps to monitor 
shifts in the wavelength scale.  The sky was approximately photometric 
throughout the night, although the seeing, as judged by the spatial profiles 
of stars, was mediocre (full width at half-maximum $\sim$1\farcs2--1\farcs8). 

Data reduction --- including bias subtraction, flatfield correction, removal of 
cosmic rays, extraction of one-dimensional spectra, and wavelength 
calibration --- closely followed standard procedures for echelle spectroscopy, 
as described by Ho \& Filippenko (1995), and will not be repeated.  We mention 
here only a few noteworthy exceptions particular to this data set.  (1) The 
short-wavelength orders of the optical setting have very little signal as 
a consequence of the low throughput of the red 
collimator and the decreasing sensitivity of the CCD at these wavelengths.  
To increase the counts in the flatfield frames at blue wavelengths without 
saturating the chip at the red end, a set of relatively long internal quartz 
lamp exposures were taken with the blue-sensitive BG12 filter.  These frames 
were combined with the usual flatfields taken with the red-sensitive NG3
filter to create a ``master'' flatfield, which has relatively uniform counts in 
the entire spectral range.  (2) In all cases, the spectra analyzed in this 
paper refer to a fixed effective aperture of 1\farcs15 $\times$ 2\farcs05.  
We experimented with the alternative optimal-extraction technique described 
by Horne (1986), but the improvement in signal-to-noise ratio (S/N) was not 
appreciable, and so we chose to retain the fixed-aperture spectra.  (3) 
The relatively short length of the slit, the presence of extended, nonuniform 
nebular emission surrounding each cluster, and the variation of the background
as the slit rotates during each exposure render ``sky subtraction'' very 
difficult in the vicinity of bright emission lines.  In almost all cases, 
the background under the object of interest differs appreciably from 
that of the neighboring pixels.  In fact, all three star clusters 
sit in local minima in the distribution of ionized gas (NGC~1569-A, 
Waller 1991; NGC 1705-1, Meurer \etal 1992; NGC~4214-1, Sargent \& Filippenko 
1991), and background subtraction introduces spurious negative features 
in the object spectra.  Although the stellar absorption features of primary 
interest to us do not have emission-line counterparts, and thus are not 
affected by this problem, we decided not to perform sky subtraction for the 
regions of the spectra containing prominent emission lines.  The exact pixel 
ranges to flag were determined by careful visual examination of each echelle 
order in the two-dimensional data frame.  This allows us, for instance, to 
obtain a more reliable representation of the combined absorption plus emission 
profiles of the main emission lines.  For the remaining portions of the 
spectrum, the background spectrum was extracted by averaging two nearby regions ($r$ = 1\farcs6 to 2\farcs5 from the cluster centroid) on opposite sides of the 
object.  (4) The dispersion in the wavelength solution of the present data 
is $\sim$0.006 and 0.007 \AA\ (1$\sigma$) for the optical and NIR settings, 
respectively, somewhat worse than that obtained by Ho \& Filippenko (1995).

At the time of the observations, HIRES was not yet equipped with an image
rotator or an atmospheric dispersion compensator.  Since the position angle of
the spectrograph slit was generally not at the parallactic angle, light losses 
from atmospheric dispersion are expected to affect observations acquired at 
high airmasses, especially for short wavelengths (Filippenko 1982).  
Specifically, an object centered on the slit as viewed on the television 
screen in fact is progressively {\it offset} from the center of the slit 
for wavelengths far from the peak of the bandpass sensitivity 
($\sim$5000--6500 \AA) of the camera.  For large airmasses, therefore, portions 
of the spectrum much bluer or redder than the peak cannot be trusted, 
especially if there are other sources of light in the vicinity of the object.  
At a declination of +65\deg, the airmass during the observation of NGC~1569-A 
remained moderately high (1.5--1.9).  NGC~1705-1 (declination $\approx$ 
--53\deg) was affected most severely, since the airmass never decreased much 
below $\sim$3.5.  NGC~4214-1, on the other hand, should be hardly affected 
(airmass \lax 1.2).  

\section{Results and Analysis}

\subsection{The Spectrum of NGC~1705-1}

A number of different photometric indicators, summarized by Marlowe \etal 
(1995), suggest that the age of NGC~1705-1 is on the order of 10--20 Myr.  
Population synthesis models (e.g., Charlot \& Bruzual 1991; Bruzual 
\& Charlot 1993) for a cluster of this age predict that the integrated light 
blueward of the $V$ band is still dominated by stars on the main sequence, 
mainly of early-B spectral type, whereas the flux at and 
redward of the $V$ band comes largely from supergiants, the majority of which 
should be of late type (O'Connell 1996).  These predictions appear 
to be in accord with the observed spectral properties of the cluster.
According to Melnick \etal (1985a), the low-resolution optical spectrum of 
NGC~1705-1 at wavelengths shorter than $\sim$6000 \AA\ is dominated by 
stars of spectral type B3~V; they mention the possible detection of 
some metal lines from later spectral types, but this was difficult to discern 
with certainty given the quality of their data.  Meurer \etal (1992) also 
published a low-resolution optical spectrum, but it yielded no new 
information because neither the wavelength coverage nor the S/N was better 
than that of the Melnick \etal data.  Lamb \etal (1985) interpreted the 
Si~IV and C~IV absorption lines seen in the UV spectrum as signatures of early 
B stars (but see York \etal 1990).  A sizable population of red supergiants 
has been invoked to explain the near-IR colors of the cluster (Melnick, Moles, 
\& Terlevich 1985; Meurer \etal 1992; Quillen \etal 1995); this seems 
justified given the presence of the Ca~II triplet absorption lines in its 
near-IR spectrum (Heckman 1996).  Melnick \etal (1985b) also detected the 
2.2 \micron\ CO absorption band head, a signature of a substantial 
population of red supergiants.

The composite nature of the NGC~1705-1 spectrum also becomes quite clear in 
our high-dispersion data.  The short-wavelength orders, despite having very 
low counts as a result of light losses caused by atmospheric dispersion 
(\S\ 2), show extremely broad Balmer and He~I absorption lines arising 
from B-type main sequence stars, as discussed by Melnick \etal (1985a).  
By contrast, a plethora of weak metal and molecular lines litter the spectrum 
at wavelengths longward of $\sim$4800 \AA; a few representative orders are 
shown in Figure 1.  Figures 2 and 3 display several additional orders of the 
cluster spectrum, juxtaposing it with the spectrum of a 
B3~V star (HR~3454) and a K5--M0 Iab-Ib star (HR~2289, or 46 $\psi^{\prime}$ 
Aur).  The B dwarf is essentially featureless at these wavelengths, while the 
red supergiant matches the cluster spectrum very closely.  The most obvious 
difference with the red supergiant spectrum is that the equivalent widths of 
the absorption lines in the cluster are much smaller, as would be expected 
from dilution by the continuum of the hotter stars.  

The weakness of the absorption lines in the spectrum of NGC~1705-1 might give 
concern that the features attributed to the cluster may be an artifact of 
imperfect background subtraction.  Such an error could occur, for instance, 
if there is a strong gradient in the intensity of the underlying light and/or 
the stellar population, so that the ``sky'' spectrum subtracted from the 
``object'' spectrum is a poor representation of the background within the 
extraction aperture.  
This, however, is unlikely to be a serious effect in the present case.  First, 
the region chosen to represent the background is located $\sim$2\asec\ away 
from the cluster (\S\ 2), which corresponds to just 50 pc at a distance of 
5 Mpc (O'Connell \etal 1994); neither the luminosity distribution nor the 
stellar population should change significantly on such a small scale.  
Second, the background level within the extraction aperture is on average 
about 3 times fainter than the cluster itself, and the equivalent widths 
of the absorption lines in the background spectrum are even smaller than those 
of the cluster spectrum (bottom of Fig. 4)\footnote{For completeness, 
we also illustrate the background spectrum of NGC~1569-A (top of Fig. 4), 
which was not included in Paper I.}.

\subsection{Velocity Dispersion and Dynamical Mass of NGC~1705-1}

It is not our immediate goal to study in quantitative terms the 
stellar content of NGC~1705-1; rather, we merely wish to establish the 
feasibility of measuring the velocity dispersion given its probable 
composition.  In the past, standard techniques for calculating velocity 
dispersions from spatially integrated spectra have been applied almost 
exclusively to relatively old ($>$ 1 Gyr) stellar systems, where red giants 
can generally serve as velocity templates.  The integrated visual light of a 
stellar population as young as 10--20 Myr, on the other hand, arises from a 
combination of hot main-sequence stars and supergiants of both high and low 
effective temperatures (\S\ 3.1).  As we argued in relation to 
NGC~1569-A (Paper I), it is still possible to obtain velocity dispersions so 
long as one restricts the analysis to portions of the spectrum whose metal 
absorption lines originate from cool supergiants.  The absorption lines from 
main-sequence stars and hot supergiants cannot be used to measure velocity 
dispersions because their line profiles are substantially broadened by a 
variety of mechanisms (pressure, rotation, microturbulence, 
macroturbulence).  Although the atmospheres of cool supergiants are still 
affected by a non-negligible amount of macroturbulence, it appears that their 
intrinsic line widths ($\sigma$) have a fairly small spread ($\sim$1.5 \kms) 
about a mean of $\sim$9 \kms\ and rarely exceed 10 \kms\ 
(for types F5--K5; Gray \& Toner 1987).  In principle, therefore, it should 
be possible to use cool supergiant stars as templates to recover velocity 
dispersions that are larger than the spread of intrinsic widths.

As in Paper I, we measured the velocity dispersion of NGC~1705-1 using the 
cross-correlation technique of Tonry \& Davis (1979), as implemented in 
the Smithsonian Astrophysical Observatory's version of 
the IRAF\footnote{IRAF is distributed by the National Optical Astronomy
Observatories, which are operated by the Association of Universities for
Research in Astronomy, Inc., under cooperative agreement with the National
Science Foundation.} task XCOR.  The Tonry \& Davis method assumes that 
the cluster spectrum can be represented as the convolution of a star spectrum 
(a template whose spectral type roughly matches the effective stellar 
population of the cluster) with a line-broadening function (taken to be 
a Gaussian) describing the line-of-sight velocity dispersion of the stars in 
the cluster. The cross-correlation function between the cluster and 
template spectrum peaks at the relative radial velocity of the two 
objects, and the width of the main velocity peak is related to the velocity 
dispersion of the cluster.  The correspondence between the width of the 
peak and the actual velocity dispersion is determined empirically by 
artificially broadening the template spectra with Gaussians of various 
dispersions and subsequently remeasuring them using the cross-correlation 
algorithm.

We restricted our analysis to the 
region between 5000 and 6280 \AA\ for three reasons: (1) it has the 
highest S/N, (2) it is unaffected by centering problems due to atmospheric 
dispersion (\S\ 2), and (3) the absorption lines come mainly from cool 
supergiants and are thus suitable for measuring velocity dispersions 
(see above).  There are 14 echelle orders in this spectral region, 
and they were processed independently using 3 template stars: HR~3422 (G8 IV), 
HR~4517 (M1 III), and HR~2289 (K5--M0 Iab-Ib).  Although the K5--M0 supergiant 
visually seems to provide an excellent match to NGC~1705-1 (Fig. 2 and 3), we 
tried the other two stars both as a consistency check and to evaluate the 
procedure adopted in Paper I for NGC~1569-A.  In the case of NGC~1569-A, the 
cluster spectrum matches better with the G8 subgiant HR~3422 than it does with 
the K5--M0 supergiant, and the subgiant was used as the template; an 
{\it a posteriori} correction had to be applied to the cluster's velocity 
dispersion  to account for the broadening contribution due to supergiants.  

Two of the orders are illustrated in Figure 5; the top panels show the 
NGC~1705-1 spectrum, the middle panels show the adopted K5--M0 supergiant 
template, and the resulting cross-correlation functions are given on the 
bottom.  All 14 orders yielded consistent values for the radial velocity (with 
respect to the template stars) and velocity dispersion, but the three 
stars gave systematically different results that are worth noting.  
The velocity dispersion obtained from the K5--M0 supergiant, which we 
assume to be the most faithful representation of the actual stellar 
content of the cluster, is $\sigma_*$ = 11.4$\pm$1.5 \kms, where the uncertainty
denotes the standard deviation of the 14 values about the mean.  The 
corresponding results for the M1~III and G8~IV templates are $\sigma_*$ = 
14.3$\pm$1.9 and 14.7$\pm$1.9 \kms, respectively; not surprisingly, they 
are larger than the dispersion derived using the supergiant, and by 
approximately the correct amount.  For instance, cross correlating 
HR~2289 with HR~4517 reveals that the supergiant is broadened 
by $\sim$7.5 \kms\ with respect to the giant, which is roughly the difference 
(subtracted in quadrature) between the cluster dispersions obtained with the 
two stars.  In the subsequent discussion, we will assume that NGC~1705-1 has 
a line-of-sight velocity dispersion of $\sigma_*$ = 11.4$\pm$1.5 \kms.

The conventional method of calculating dynamical masses of globular clusters 
(e.g., Illingworth 1976) makes use of parameters obtained from surface 
photometry and the core velocity dispersion.  Unfortunately, detailed surface 
photometry of our distant clusters is not yet available, and so we resort 
to a cruder estimate of the mass using the virial theorem.  For a spherically 
symmetric, gravitationally bound system of stars having equal masses and an 
isotropic velocity distribution [$\sigma^2$(total) = 3$\sigma_*^2$], the 
dynamical mass is given by $M\,=\,3\sigma_*^2 R/G$, where $R$ is the effective 
gravitational radius. The condition of spherical symmetry probably holds, 
since NGC~1705-1 does not exhibit any obvious substructure and appears 
essentially pointlike (O'Connell \etal 1994).  
We take $R$ = $R_{\rm h}$, the half-light radius.  
In the context of other probable sources of uncertainties contained in our
simplistic application of the virial theorem, the observationally 
straightforward half-light radius suffices for the present purposes.  
O'Connell \etal (1994) find $R_{\rm h}\, <$ 3.4 pc for NGC~1705-1, assuming a 
(poorly known) galaxy distance of 5 Mpc.  Meurer \etal (1995) reexamined the 
{\it HST} images of O'Connell \etal using a different method of analysis, and 
they found a substantially smaller size of $R_{\rm h}$ = 0.9$\pm$0.2 pc 
(adjusted to the distance adopted by O'Connell et al.).  Using Meurer et al.'s 
revised size, we derive a mass of (8.2$\pm$2.1)\e{4} \solmass\ from the 
virial equation.  The implied mass density of the cluster is 2.7\e{4} 
\solmass\ pc$^{-3}$.

As discussed in Paper I, relaxing some of the assumptions made in our 
application of the virial equation most likely will {\it increase} the final 
mass.  It would be completely inappropriate to apply the virial theorem, of 
course, if the cluster were not bound; nevertheless, it is difficult to 
imagine how this could be.  Because of the small size of the cluster, any 
conceivable form of perturbation that might disperse the stars will traverse 
the cluster on a timescale shorter than 1 Myr, whereas the age of the cluster 
is much longer, on the order of 10--20 Myr.  Although difficult to prove, a 
similar heuristic argument has been made by others (Whitmore \etal 1993; 
Whitmore \& Schweizer 1995; Maoz \etal 1996).  On the other hand, it is 
unlikely that such a young cluster, if it truly has a mass comparable to that 
of globular clusters, has had time to completely come to virial equilibrium.  
A typical globular cluster takes about 10$^9$ yr to relax dynamically (Binney 
\& Tremaine 1987).  Since the final velocity dispersion of a virialized 
cluster will be higher than its present value, our derived mass probably has 
been underestimated.  The probable presence of mass segregation also 
potentially biases the velocity dispersion to lower values.   Mass segregation 
for the more massive stars appears to have taken place even for a cluster as 
young as R136 (3--4 Myr; Hunter \etal 1996) in the center of 30 Doradus 
(Brandl \etal 1996; but see Hunter \etal 1996); thus, it seems plausible that 
older clusters such as NGC~1569-A and NGC~1705-1 should be similarly affected.
The integrated spectrum predominantly samples the most luminous, massive 
stars; if mass segregation has set in, these stars will have sunk toward the 
center of the cluster, and their velocity dispersion will be lower than that 
of the entire system.   The quoted mass for NGC~1705-1, therefore, can be 
regarded as a fairly safe lower limit, and the true mass can certainly be as 
high as 10$^5$ \solmass\ or more. Not knowing the distance to NGC~1705 with 
better accuracy certainly propagates additional uncertainty into the mass 
estimate, but note that the physical size (computed from the angular size and 
the distance) only enters linearly into the virial equation.

Since our approach for estimating the dynamical mass is more primitive than 
the conventional method, it is worth considering how large an error we could 
have made.  The compilation of Mandushev \etal (1991) conveniently tabulates 
globular cluster masses derived using Illingworth's (1976) method, as well as 
core velocity dispersions and half-light radii.  Given that the total (spatially
integrated) velocity dispersion is not much smaller than the core velocity 
dispersion (Illingworth finds a difference of only a few percent between the 
two), we can compute the masses using our adopted formula and compare them to 
the values in Mandushev et al.  We find that in general there is fairly 
close agreement between the masses obtained from the two methods.  The 
discrepancies rarely exceed a factor of two, and this level of uncertainty 
does not change our overall conclusions.  

Several previous studies have attempted to estimate the stellar mass in
NGC~1705-1 using stellar population models.  However, since most of the 
observed properties of the cluster sample only the massive stars, obtaining 
the total cluster mass using these models necessarily must extrapolate a 
poorly constrained stellar initial mass function (IMF) to lower, unobserved 
stellar masses.  Masses obtained by such indirect, model-dependent methods are 
notoriously difficult to assess.  The dynamical mass we have derived is in 
substantial disagreement with the mass given by Melnick \etal (1985a); even 
after accounting for the larger distance they adopted (8.7 Mpc), their quoted 
mass of 7\e{6} \solmass\ still exceeds ours by about a factor of 50.  
Similarly, Meurer \etal (1992) used a different set of models and derived 
$M$ = 1.5\e{6} \solmass, close to the value of Melnick \etal after adjusting 
to a common distance (Meurer \etal assumed 4.7 Mpc).  The most recent estimate 
comes from the {\it HST} UV imaging study of Meurer \etal (1995), who, from 
the luminosity at 2200 \AA\ (assuming a distance of 6 Mpc), deduced $M$ = 
2.5\e{5} \solmass; this is in acceptable agreement with our dynamical 
measurement, especially considering that our value may be an underestimate.

\subsection{The Spectrum of NGC~4214-1}

The HIRES spectrum of NGC~4214-1 is dominated by a featureless continuum
from hot stars on which are imprinted broad Balmer (H$\delta$ to H\bet) and
He~I (\lamb\lamb 4026, 4121, 4144, 4388, 4472, 4713, and 4922) absorption
lines.  Ca~II H \& K and Na~I D absorption are present (Fig. 6{\it a}), but 
these lines arise from the interstellar medium of the Galaxy and of NGC~4214.
Unfortunately, very few metal lines from stars are seen.  The few that can be
tentatively identified include a blend of Si~IV \lamb4088.9 and O~II
\lamb4089.3, N~III \lamb4200, Mg~II \lamb4481, and a complex near 4650 \AA\
possibly due to C~II, C~III, and O~II.  These lines probably come from a
mixture of B dwarfs and supergiants (e.g., Kilian, Montenbruck, \& Nissen
1991; Lennon, Dufton, \& Fitzsimmons 1992), and cannot be used to determine
velocity dispersions (\S\ 3.2).  Given that the age of the cluster is only
4--5 Myr (Leitherer \etal 1996), the absence of stars of later spectral types
is not surprising, if all the stars formed in a single burst.  

The emission lines included in our extraction aperture show complex velocity 
structure (Fig. 6{\it b}).  Nearly all the strong lines exhibit a 
double-peaked profile, with a very weak third peak visible in some cases.
The radial velocity of the stronger 
of the two components agrees closely with the systemic velocity of the 
galaxy (291 \kms; de Vaucouleurs \etal 1991), while the secondary peak is 
redshifted by approximately 75--80 \kms.  Both components have broad, 
extended bases and cannot be fitted adequately with a single Gaussian function.
The H\al\ line, in particular, contains obvious broad wings with a full 
width near zero intensity (FWZI) of $\sim$760 \kms;\footnote{Sargent \& 
Filippenko (1991) and Marlowe \etal (1995) also noted the presence of broad 
H\al\ in NGC~4214; Marlowe \etal estimated FWZI $\approx$ 600 \kms.}similarly,
He~I \lamb5876 (Fig. 6{\it a}) has broad wings with FWZI $\approx$ 600 \kms.
 
\section{Implications for the Formation of Globular Clusters}

The large mass and mass density of NGC~1705-1 lend credence to the popular idea 
that some compact, luminous, young clusters may be recently formed globular 
clusters.  The total stellar mass of NGC~1705-1 (8.2\e{4} \solmass) coincides 
with the median mass of evolved, Galactic globular clusters (8.1\e{4} 
\solmass; Mandushev \etal 1991).  The cluster half-light radius (0.9 pc), 
if anything, is even smaller than that of typical globular clusters; the 
median $R_{\rm h}$ = 2.6 and 4.1 pc in the samples of van den Bergh, Morbey, 
\& Pazder (1991) and Mandushev \etal (1991), respectively.  Dynamical 
evolution of the cluster will cause it to expand slightly as it ages (Elson 
1992), however, and the final cluster radius may turn out to be normal.  
O'Connell \etal (1994) determined an integrated absolute visual magnitude of 
--14.0 for NGC~1705-1.  Since a 10--20 Myr old cluster is expected to dim by 6 
to 7 mag in the $V$ band after 10--15 Gyr (e.g., Bruzual \& Charlot 1993), 
NGC~1705-1 will fade to $M_V\,\approx$ --7 to --8 mag in its maturity, in 
excellent agreement with the peak of the nearly universal luminosity function 
of globular cluster systems ($\langle M_V \rangle\,\approx$ --7.3 mag; Harris
1991).  Moreover, if mass loss during advanced stages of stellar evolution 
does not significantly reduce the mass of the cluster, NGC~1705-1 will 
attain a mass-to-light ratio of 0.7--1.6 ($M/L_V$)$_{\odot}$, again 
fully consistent with the observed, narrow range for Galactic globular 
clusters [0.7--2.9 ($M/L_V$)$_{\odot}$; Mandushev \etal 1991].

The properties of NGC~1705-1 agree remarkably well with those of NGC~1569-A 
(see Table 1).  NGC~1569-A has a comparable age (10--20 Myr), is just as 
luminous ($M_V$ = --14.1 mag), and is only slightly less compact ($R_{\rm h}$
= 1.9 pc).  Ho \& Filippenko (1996) measured a velocity dispersion of 
$\sigma_*$ = 15.7$\pm$1.5 \kms, from which they derived a cluster mass of 
(3.3$\pm$0.5)\e{5} \solmass, a mass density of 1.1\e{4} \solmass\ pc$^{-3}$, 
and a predicted mass-to-light ratio of 2.5--6.3 ($M/L_V$)$_{\odot}$ after 
10--15 Gyr.  Again, all of these properties strongly resemble those of the 
majority of evolved globular clusters in the Galaxy.

It would be premature to conclude that the above results apply to all of the 
compact, super star clusters discovered with {\it HST}; such an extrapolation,
based on just two objects, would require a sizable leap of faith.  Furthermore, 
the environments of the two clusters studied here are quite different from 
those of the other galaxies in which super star clusters have been found.  Both 
NGC~1569 and NGC~1705 are amorphous galaxies (Sandage \& Brucato 1979),
distinguished by their relatively smooth, centrally-concentrated light 
distribution that resembles that of elliptical and S0 galaxies, while at the 
same time sharing many of the global characteristics (e.g., metallicity, 
average colors, star-formation rate) of Magellanic irregular 
galaxies (Gallagher \& Hunter 1987).  NGC~1569 has two luminous clusters, 
and NGC~1705 just one, both centrally located and far brighter than the
other clusters known to be present in the same galaxies.  Nevertheless, 
the mere fact that we are witnessing the recent birth of \gax 10$^5$ solar 
masses of stars packed into a volume $\sim$3--30 pc$^{3}$ is notable, and 
although the exact details of the formation of these young clusters may differ 
in some respects from those of globular clusters born during the era of galaxy 
formation, studying them should still prove to be illuminating.  For instance, 
clusters such as NGC~1569-A and NGC~1705-1 evidently can form from an 
environment far more metal rich than that which produced the first generation 
of globular clusters: [O/H] = --0.6 for NGC~1569-A (Heckman \etal 1995), and 
[O/H] = --0.35 for NGC~1705-1 (Marlowe \etal 1995), whereas the metallicity 
distribution of globular clusters in the Galactic halo peaks at 
$\langle \rm{[Fe/H]} \rangle\,\approx$ --1.6 (Armandroff \& Zinn 1988).
Models of globular cluster formation relying principally on the physical 
conditions of extremely metal-poor gas (e.g., Murray \& Lin 1989, 1993; 
Richtler 1993) evidently do not apply to these young clusters.  Perhaps 
they are more closely related to the population of metal-rich {\it disk} 
globular clusters, whose $\langle \rm{[Fe/H]} \rangle\,\approx$ --0.5 
(Armandroff \& Zinn 1988).

In the spirit of the Searle \& Zinn (1978) picture for the formation of the 
Galactic halo in which the halo is assembled from the merging of a number of 
protogalactic fragments, it has been suggested that the halo 
globular clusters perhaps represent the ``nuclei'' of accreted, stripped dwarf 
satellites (Zinnecker \etal 1988; Larson 1990, and references therein).
Our finding that the central star clusters in NGC~1569 and NGC~1705 appear to 
be young globular clusters lends some observational support for this scenario.

As revealed by recent high-resolution observations from {\it HST} (see \S\ 1 
and additional references in Ho 1996), cluster formation 
appears to be a very common, if not the dominant, mode of star formation 
in a wide range of starburst environments.  The brightest members of the 
cluster population in these systems closely resemble NGC~1569-A and NGC~1705-1 
in terms of physical dimension and luminosity (and presumably mass).  The 
inner few hundred parsecs of the famous starburst galaxy M82, for instance, 
harbor a swarm of over 100 such clusters (O'Connell \etal 1995), and 
undoubtedly many more remain hidden by dust from direct view.  
Super star clusters, therefore, may constitute the basic building blocks 
or ``cells'' of star formation in starbursts, and intensive observational 
dissection of clusters such as NGC~1569-A and NGC~1705-1 should prove to 
be a valuable tool for understanding the starburst phenomenon in general.

As an illustration of the potential application of these clusters, consider 
two simple examples.  First, the apparent similarity between the stellar 
content of
super star clusters and globular clusters, as implied by the mass-to-light 
ratios we deduced, may have important consequences for the IMF of starbursts.  
Since we have shown that at least {\it some} super star clusters seem to be 
genuine, young globular clusters, and the latter are known to contain many 
low-mass stars, the implication is that low-mass stars are also formed in 
starburst regions.  In the well-known case of the center of M82, it has been 
suggested that the IMF is biased toward high-mass stars (Rieke \etal 1980, 
1993).  This result, while provocative, remains controversial (e.g., Scalo 
1986, 1987; Zinnecker 1996), and determination of the dynamical masses for 
a larger sample of clusters may have a strong bearing on this issue.

Second, the existence of massive, bound star clusters puts some constraints on 
the physical state of the dense interstellar medium from which they formed.  
The efficiency of star formation (rate of conversion from gas to stars) in 
the disk of the Galaxy is fairly low;  Myers \etal (1986) estimate it to be 
on the order of only a few percent, with large variations from region to 
region.  For newly formed star clusters to be gravitationally bound, however, 
the efficiency of star formation must be much higher, 
perhaps 20\%--50\% (Hills 1980; Elmegreen 1983; Lada, Margulis, \& Dearborn 
1984).  A sufficiently large amount of gas needs to be converted 
into stars in order for the cluster to acquire enough gravitational binding 
energy.  If super star clusters typically attain masses of $\sim$10$^5$ 
\solmass, their progenitor molecular clouds must have masses ranging 
from 2\e{5} to 5\e{5} \solmass; considering the compactness of the 
clusters, the ``protocluster cores'' must also be fairly dense, denser than 
typical giant molecular clouds.  Such a population of clouds may indeed 
be present in the starburst region of M82.  Both 
Brouillet \& Schilke (1993) and Shen \& Lo (1995) have published mm-wave 
interferometric maps of the dense molecular gas tracer HCN, and they find 
a large number of massive clouds located in the central few hundred parsecs 
of the galaxy.  The median cloud listed in Table 2 of Brouillet \& Schilke, 
for example, has a radius of 50 pc and a mass of 10$^6$ \solmass.  It is 
tempting to identify these massive, dense clouds as the progenitors of the 
super star clusters in M82.

\section{Summary}

Together with the results reported in Paper I, we have 
measured the stellar velocity dispersion of two luminous, compact star 
clusters (NGC~1569-A and NGC~1705-1) by analyzing high-dispersion optical 
spectra of their integrated light.  The clusters are relatively young, with 
ages of 10--20 Myr.  Both clusters were recently imaged with 
{\it HST} by O'Connell \etal (1994), and hence their sizes are known.  If they 
are gravitationally bound, application of the virial theorem implies 
that the clusters in NGC~1569 and NGC~1705 have stellar masses of 
(3.3$\pm$0.5)\e{5} and (8.2$\pm$2.1)\e{4} \solmass, and mass densities of 
1.1\e{4} \solmass\ pc$^{-3}$ and 2.7\e{4} \solmass\ pc$^{-3}$, respectively.  
The masses, mass densities, and the predicted mass-to-light ratios (after 
aging by 10--15 Gyr) of these objects are very similar to those of evolved 
globular clusters in the Galaxy.  NGC~1569-A and NGC~1705-1, therefore, appear 
to be genuine, {\it young} globular clusters.  Since the most luminous 
members in other cluster systems discovered with {\it HST} have photometric 
properties resembling those of the two clusters studied here, we suggest that 
they too may be similar in nature.  The gas clouds from which the 
clusters formed must also have been very massive and compact, and we 
tentatively identify the dense molecular clouds recently found in the 
center of M82 as possible progenitors.

\acknowledgments

The W. M. Keck Observatory, made possible by the generous and visionary
gift of the W.~M. Keck Foundation, is operated as a scientific partnership
between the California Institute of Technology and the University of
California.  We thank Tom Bida, Meg Whittle, and Joel Aycock for 
technical support during the observations, and Aaron Barth for helping with 
some of the preliminary preparations.  We are grateful to Jay Anderson for 
generously making available his software for removing cosmic rays.  This 
work benefited from correspondence with Jay Anderson, Aaron Barth, 
Eduardo Bica, Lewis Jones, Fred Lo, Bob O'Connell, and John Stauffer.  Some 
recommendations made by an anonymous referee improved the final version of 
the paper.  The research of L.~C.~H. is funded by a postdoctoral fellowship
from the Harvard-Smithsonian Center for Astrophysics, while A.~V.~F. receives
financial support from the National Science Foundation (grant AST--9417213) 
and NASA (grant AR--05792.01-94A from the Space Telescope Science Institute).  
Partial travel support was provided by the California
Association for Research in Astronomy.  During the course of this work,
A.~V.~F. held an appointment as a Miller Research Professor in the Miller
Institute for Basic Research in Science (U.~C. Berkeley).

\clearpage

\centerline{\bf{References}}
\medskip

\refindent 
Armandroff, T.~E., \& Zinn, R. 1988, \aj, 96, 92

\refindent 
Arp, H., \& Sandage, A. 1985, AJ, 90, 1163

\refindent 
Barth, A. J., Ho, L. C., Filippenko, A. V., Gorjian, V., Malkan,
M., \& Sargent, W. L. W. 1996, in IAU Colloq. 157, Barred Galaxies, ed.\ R.
Buta, B. G. Elmegreen, \& D. A. Crocker (San Francisco: ASP), 94

\refindent 
Barth, A.~J., Ho, L.~C., Filippenko, A.~V., \& Sargent, W.~L.~W. 1995, \aj,
110, 1009

\refindent 
Benedict, G.~F., \etal 1993, \aj, 105, 1369

\refindent 
Binney, J., \& Tremaine, S. 1987, Galactic Dynamics (Princeton: Princeton
Univ. Press)


\refindent
Brandl, B., Sams, B., Bertoldi, F., Eckart, A., Genzel, R., Drapatz, S., 
Hofmann, R., Loewe, M., \& Quirrenbach, A. 1996, \apj, in press

\refindent
Brouillet, N., \& Schilke, P. 1993, \aa, 277, 381

\refindent
Bruzual A., G., \& Charlot, S. 1993, \apj, 405, 538

\refindent
Charlot, S., \& Bruzual A., G. 1991, \apj, 367, 126



\refindent 
de Vaucouleurs, G., de Vaucouleurs, A., Corwin, H.~G., Jr., Buta, R.~J.,
Paturel, G., \& Fouqu\'e, R. 1991, Third Reference Catalogue of Bright
Galaxies (New York: Springer)





\refindent
Dubath, P., Mayor, M., \& Meylan, G. 1993, in The Globular Cluster-Galaxy
Connection, ed. G.~H. Smith \& J.~P. Brodie (San Francisco: ASP), 557

\refindent
Elmegreen, B.~G. 1983, \mnras, 203, 1011

\refindent
Elson, R.~A.~W. 1992, \mnras, 256, 515


\refindent 
Filippenko, A.~V. 1982, \pasp, 94, 71


\refindent
Gallagher, J.~S., \& Hunter, D.~A. 1987, \aj, 94, 43

\refindent
Gray, D.~F., \& Toner, C.~G. 1987, \apj, 322, 360

\refindent 
Gunn, J.~E., Griffin, R.~F., Griffin, R.~E.~M., \& Zimmerman, B.~A. 1988, \aj, 
96, 198

\refindent 
Harris, W.~E. 1991, \annrev, 29, 543

\refindent 
Heckman, T.~M. 1996, private communication

\refindent 
Heckman, T.~M., Dahlem, M., Lehnert, M.~D., Fabbiano, G., Gilmore, D., \&
Waller, W.~H. 1995, \apj, 448, 98

\refindent 
Hills, J.~G. 1980, \apj, 225, 986


\refindent 
Ho, L.~C. 1996, Rev. Mex. Astr. Astrofis., in press

\refindent 
Ho, L.~C., \& Filippenko, A.~V. 1995, \apj, 444, 165 (Erratum: 1996, \apj, 
June 1 issue)

\refindent 
Ho, L.~C., \& Filippenko, A.~V. 1996, \apj, in press (Paper I)

\refindent 
Holtzman, J.~A., \etal 1992, \aj, 103, 691

\refindent 
Horne, K. 1986, \pasp, 98, 609

\refindent 
Hunter, D.~A., O'Connell, R.~W., \& Gallagher, III, J.~S. 1994, \aj, 108, 84

\refindent 
Hunter, D.~A., O'Neil, E.~J., Lynds, R., Shaya, E.~J., Groth, E.~J., \&
Holtzman, J.~A. 1996, \apj, 459, L27

\refindent 
Illingworth, G. 1976, \apj, 204, 73

\refindent 
Kilian, J., Montenbruck, O., \& Nissen, P.~E. 1991, \aas, 88, 101

\refindent 
Lada, C.~J., Margulis, M., \& Dearborn, D. 1984, \apj, 285, 141

\refindent 
Lamb, S.~A., Gallagher, J.~S., Hjellming, M.~S., \& Hunter, D.~A. 1985, \apj,
291, 63

\refindent 
Larson, R.~B. 1990, \pasp, 102, 709


\refindent 
Leitherer, C., Vacca, W.~D., Conti, P.~S., Filippenko, A.~V., Robert, C., \&
Sargent, W.~L.~W. 1996, \apj, in press

\refindent 
Lennon, D.~J., Dufton, P.~L., \& Fitzsimmons, A. 1992, \aas, 94, 569

\refindent 
Leonard, P.~J.~T., \& Merritt, D. 1989, \apj, 339, 195

\refindent 
Lutz, D. 1991, \aa, 245, 31

\refindent 
Mandushev, G., Spassova, N., \& Staneva, A. 1991, \aa, 252, 94

\refindent 
Maoz, D., Barth, A.~J., Sternberg, A., Filippenko, A.~V.,
Ho, L.~C., Macchetto, F.~D., Rix, H.-W., \& Schneider, D.~P. 1996, \aj,
in press

\refindent 
Marlowe, A.~T., Heckman, T.~M., Wyse, R.~F.~G., \& Schommer, R. 1995, \apj,
438, 563

\refindent 
Mateo, M. 1993, in The Globular Cluster-Galaxy Connection, ed. G.~H. Smith
\& J.~P. Brodie (San Francisco: ASP), 387

\refindent 
Mathieu, R.~D. 1984, \apj, 284, 643

\refindent 
Melnick, J., Moles, M., \& Terlevich, R. 1985a, \aa, 149, L24

\refindent 
Melnick, J., Terlevich, R., \& Moles, M. 1985b, \revmex, 11, 91

\refindent 
Meurer, G.~R., Freeman, K.~C., Dopita, M.~A., \& Cacciari, C. 1992, \aj, 103,
60

\refindent 
Meurer, G.~R., Heckman, T.~M., Leitherer, C., Kinney, A., Robert, C., \&
Garnett, D.~R. 1995, \aj, 110, 2665

\refindent 
Murray, S.~D., \& Lin, D.~N.~C. 1989, \apj, 339, 933

\refindent 
Murray, S.~D., \& Lin, D.~N.~C. 1993, in The Globular Cluster-Galaxy
Connection, ed. G.~H. Smith \& J.~P. Brodie (San Francisco: ASP), 738

\refindent
Myers, P.~C., Dame, T.~M., Thaddeus, P., Cohen, R.~S., Silverberg, R.~F.,
Dwek, E., \& Hauser, M.~G. 1986, \apj, 301, 398

\refindent
O'Connell, R.~W. 1996, private communication

\refindent 
O'Connell, R.~W., Gallagher, J.~S., \& Hunter, D.~A. 1994, \apj, 433, 65

\refindent 
O'Connell, R.~W., Gallagher, J.~S., Hunter, D.~A., \& Colley, W.~N. 1995, \apj,
446, L1

\refindent 
O'Connell, R.~W., \& Mangano, J.~J. 1978, \apj, 221, 62

\refindent 
Oke, J.~B., \& Gunn, J.~E. 1983, \apj, 266, 713

\refindent 
Quillen, A.~C., de Zeeuw, P.~T., Phinney, E.~S., \& Phillips, T.~G. 1992,
\apj, 391, 121

\refindent 
Richtler, T. 1993, in The Globular Cluster-Galaxy Connection,
ed. G.~H. Smith \& J.~P. Brodie (San Francisco: ASP), 375

\refindent 
Rieke, G.~H., Lebofsky, M.~J., Thompson, R.~I., Low, F.~J., \& Tokunaga, A.~T.
1980, \apj, 238, 24

\refindent 
Rieke, G.~H., Loken, K., Rieke, M.~J., \& Tamblyn, P. 1993, \apj, 412, 99

\refindent 
Sandage, A.~R., \& Brucato, R. 1979, \aj, 84, 472

\refindent 
Sargent, W.~L.~W., \& Filippenko, A.~V. 1991, \aj, 102, 107


\refindent 
Scalo, J.~M. 1986, Fundam. Cosm. Phys., 11, 1

\refindent 
Scalo, J.~M. 1987, in Starbursts and Galaxy Evolution, ed. T.~X. Thuan, T.
Montmerle, \& J.~T.~T. Van (Guf sur Yvette: Editions Fronti\'eres), 445



\refindent 
Searle, L., \& Zinn, R. 1978, \apj, 225, 357

\refindent 
Shen, J., \& Lo, K.-Y. 1995, \apj, 445, L99

\refindent 
Smith, G.~H., \& Brodie, J.~P. (eds.) 1993, The Globular Cluster-Galaxy 
Connection (San Francisco: ASP)

\refindent 
Tonry, J., \& Davis, M. 1979, \aj, 84, 1511



\refindent
van den Bergh, S., Morbey, C., \& Pazder, J. 1991, \apj, 375, 594

\refindent
Vogt, S.~S. 1994, UCO/Lick Observatory Technical Report No. 67

\refindent
Vogt, S.~S., \etal 1994, Proc. SPIE, 2198, 362

\refindent 
Waller, W.~H. 1991, \apj, 370, 144

\refindent 
Whitmore, B.~C., \& Schweizer, F. 1995, \aj, 109, 960

\refindent 
Whitmore, B.~C., Schweizer, F., Leitherer, C., Borne, K., \& Robert, C. 1993,
\aj, 106, 1354

\refindent
York, D.~G., Caulet, A., Rybski, P., Gallagher, J., Blades, J.~C., Morton,
D.~C., \& Wamsteker, W. 1990, \apj, 351, 412


\refindent
Zepf, S.~E., Carter, D., Sharples, R.~M., \& Ashman, K.~M. 1995, \apj, 445,
L19

\refindent
Zinnecker, H. 1996, in The Interplay Between Massive Star Formation, the ISM
and Galaxy Evolution, ed.  D. Kunth et al. (Paris: Editions Fronti\'eres), in
press

\refindent
Zinnecker, H., Keable, C.~J., Dunlop, J.~S., Cannon, R.~D., \& Griffiths,
W.~K. 1988, in The Harlow-Shapley Symposium on Globular Cluster Systems in
Galaxies, ed. J.~E. Grindlay \& A.~G.~D. Philip (Kluwer), 603


\clearpage
\psfig{file=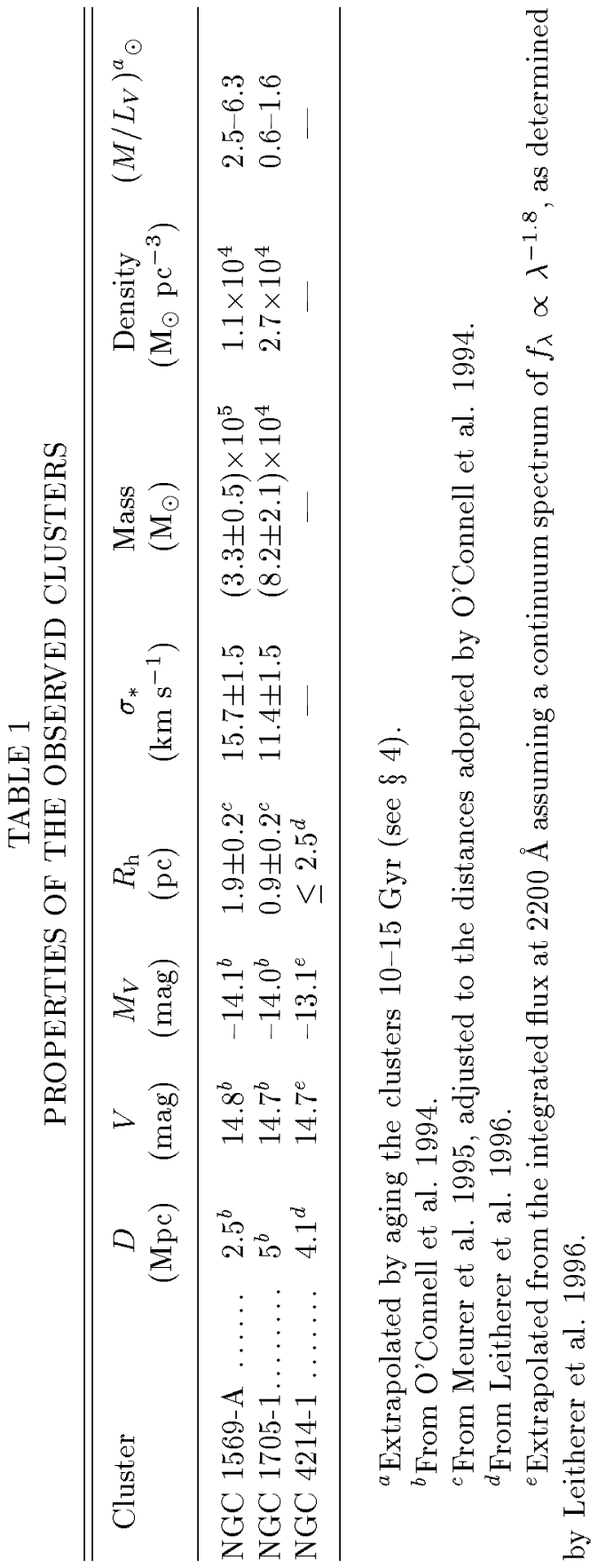,height=6.0truein,angle=180}

\clearpage

\centerline{\bf{Figure Captions}}
\medskip

Fig. 1. --- Illustration of a portion of the HIRES spectrum of NGC~1705-1 
($\sim$5100--5580 \AA).  The data were smoothed with a boxcar function 
of 3 pixels (6.4 \kms) in order to slightly improve the S/N for the sake of 
presentation, and the continuum was normalized to unity.  Numerous metal 
and molecular lines due to late-type stars are detected.  Two regions in the 
top panel (denoted by horizontal lines) were corrupted by a cosmetic flaw 
on the CCD.  The small gap in the spectrum in the middle and bottom panels 
occurs where the CCD was insufficiently large to entirely span an order.

\medskip
Fig. 2. --- Illustration of additional orders of the HIRES spectrum of 
NGC~1705-1 ($\sim$5660--5970 \AA).  The cluster data ({\it b}) have been 
slightly smoothed (see Fig. 1).  Shown for comparison in the same panel is the 
spectrum of the B3~V star HR~3454 ({\it a}), offset in the ordinate by an 
arbitrary amount for clarity.  The narrow features in the last order are due 
to telluric absorption, and the incomplete broad absorption line 
is He~I \lamb 5876.  The template star, HR~2289 (K5--M0 Iab-Ib), is shown in 
the bottom panel ({\it c}).  The two strongest absorption lines in the 
spectra shown of NGC~1705-1 and HR~2289 are Na~I~D \lamb\lamb5890, 5896.  
Gaps in the spectrum occur where the CCD was insufficiently large 
to entirely span an order.  The continuum level was normalized to unity 
and the wavelength scale shifted to the rest frame of the objects.

\medskip
Fig. 3. --- Illustration of additional orders of the HIRES spectrum of 
NGC~1705-1 ($\sim$5950--6290 \AA).  Same as in Figure 2.

\medskip
Fig. 4. --- Comparison of a typical extraction of the cluster spectra with the 
corresponding background (``sky'') spectra.  In the case of NGC~1569-A 
({\it top half}) and NGC~1705-1 ({\it bottom half}), the background spectra 
were scaled by factors of 5.1 and 2.6, respectively.  The continuum level of 
all the spectra were normalized to unity, and the individual spectra were 
lightly smoothed (see Fig. 1) and offset in the ordinate by an arbitrary 
amount for clarity.  Note that the cluster spectra in this figure have lower 
S/N than those in others because only a single exposure is shown.

\medskip
Fig. 5. --- ({\it a}) Example of the cross-correlation technique applied to
one of the orders ($\sim$5330--5410 \AA).  The {\it top} panel shows the 
spectrum of NGC~1705-1 and the {\it middle} panel the template star HR~2289 
(K5--M0 Iab-Ib), both normalized to unity and shifted to their rest frame.  The 
cluster spectrum has been slightly smoothed (see Fig. 1).
The {\it bottom} panel plots the cross-correlation function 
(CCF) between the cluster and the star.  The width of the main velocity peak 
of the CCF is related to the velocity dispersion of the object.
In this order, the S/N per pixel of the continuum in the cluster
(before smoothing) ranges from 33 at the blue end to 55 at the red end.  The
corresponding values for the template star are S/N = 300 to 500.  ({\it b}) 
Same as ({\it a}), but for another order ($\sim$6065--6155 \AA).  The S/N is 
38--53 for NGC~1705-1 and 450--700 for the template.

\medskip
Fig. 6. --- Sample spectra of NGC~4214-1.  The spectra were lightly smoothed 
(see Fig. 1), shifted to the rest frame of the galaxy, and normalized so that 
the continuum level is unity.  Portions of two orders from the optical setting 
are shown in ({\it a}).  The top panel displays the Ca~II H \& K interstellar 
absorption lines arising from the Galaxy and from NGC~4214.  The Ca~II H line 
from NGC~4214 is partially contaminated by [Ne~III] \lamb3968 emission and 
H$\epsilon$ emission and absorption.  The bottom panel shows the region 
containing the Na~I D2 (\lamb5890) and D1 (\lamb5896) interstellar absorption 
lines; the components arising from the Galaxy and NGC~4214 are indicated. 
[Note that the apparently anomalous radial velocities of the Galactic 
components of the interstellar lines result from shifting the spectra to the 
rest frame of NGC~4214, whose systemic (heliocentric) velocity is assumed to 
be 291 \kms\ (de Vaucouleurs \etal 1991).]  The double-peaked emission line 
is He~I \lamb5876, superposed on a broad base (FWZI $\approx$ 600 \kms).  
Portions of some orders containing strong emission lines 
are plotted in ({\it b}); the spectra in the two left panels were taken 
with the optical setting, and the ones in the right panels 
with the NIR setting.  All the lines contain at least two clear
emission peaks.  The slight depression in the continuum surrounding the 
H\bet\ emission line is caused by H\bet\ absorption.  The H\al\ emission 
line has broad H\al\ wings with FWZI $\approx$ 760 \kms.

\clearpage
\begin{figure}
\figurenum{1}
\plotone{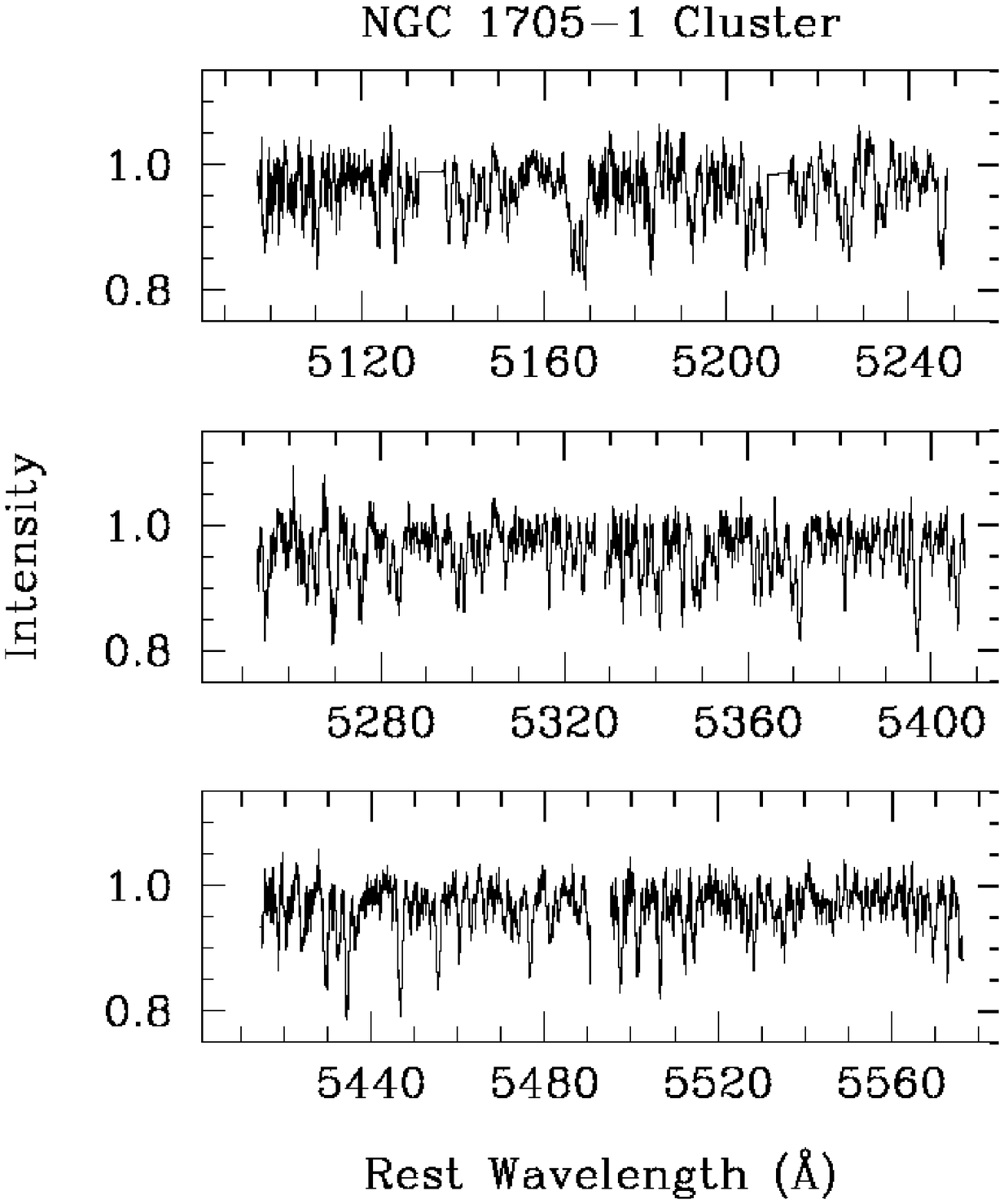}
\caption{}
\end{figure} 
 
\clearpage
\begin{figure}
\figurenum{2}
\plotone{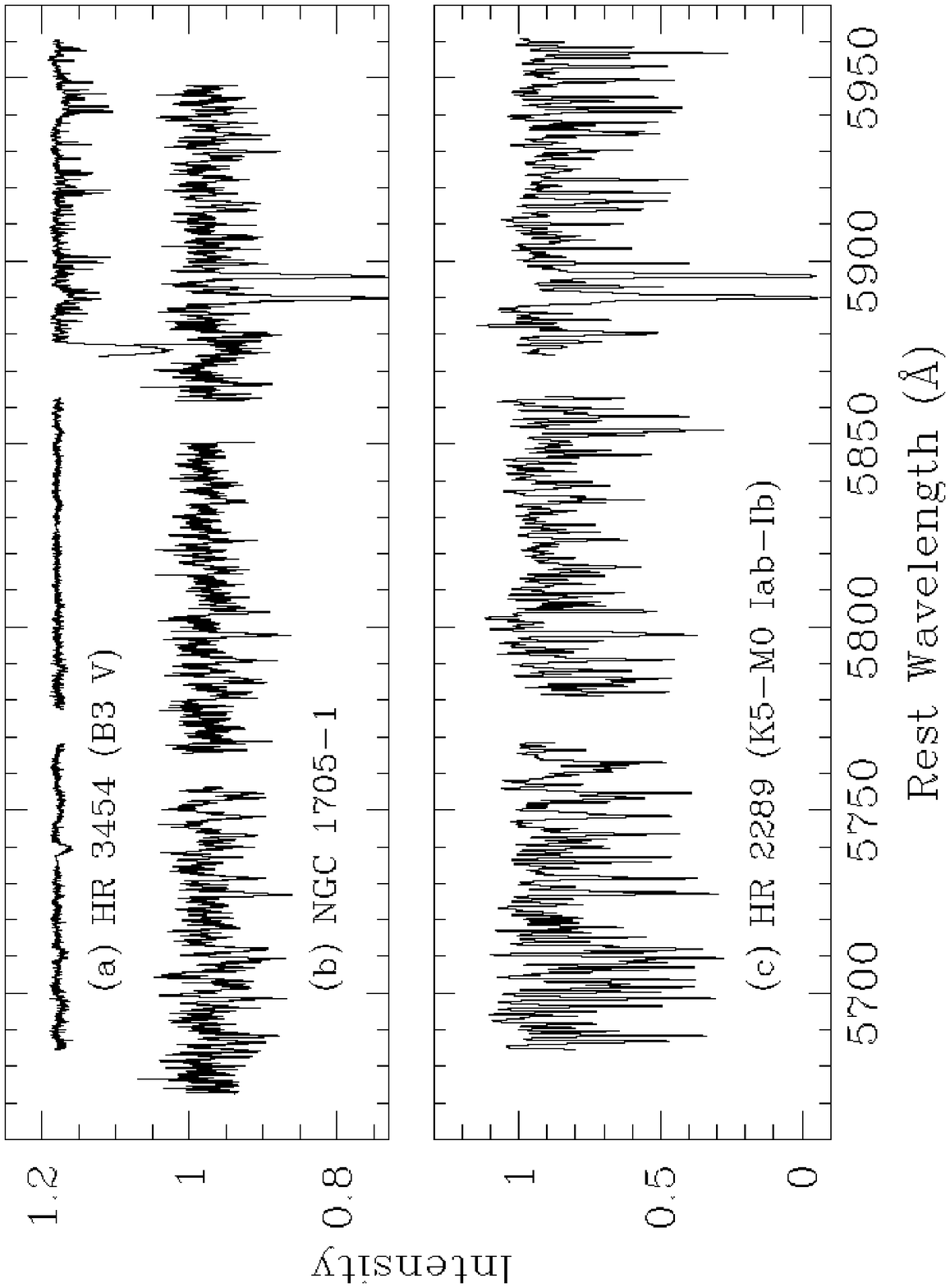}
\caption{}
\end{figure} 
 
\clearpage
\begin{figure}
\figurenum{3}
\plotone{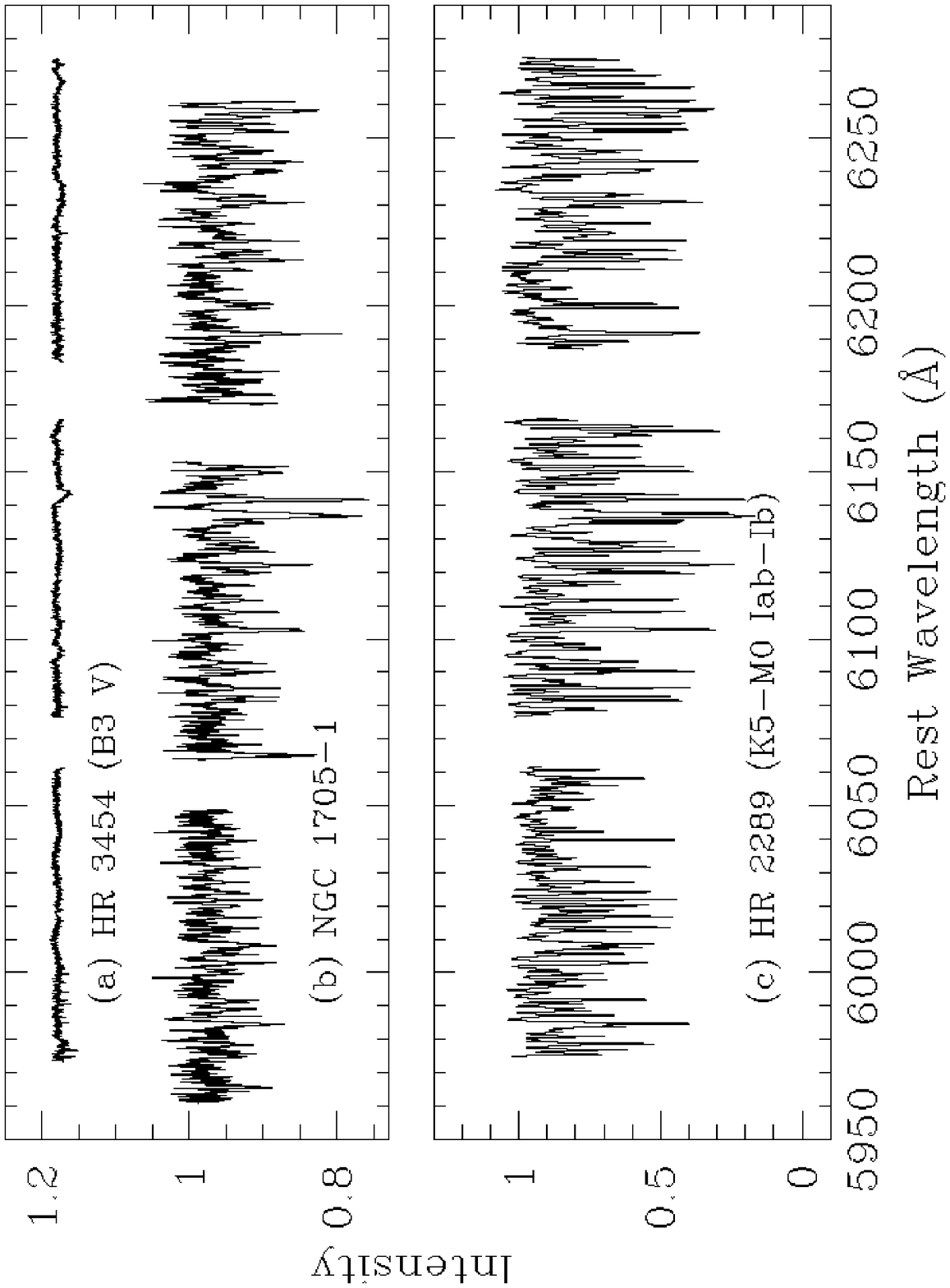}
\caption{}
\end{figure}

\clearpage
\begin{figure}
\figurenum{4}
\plotone{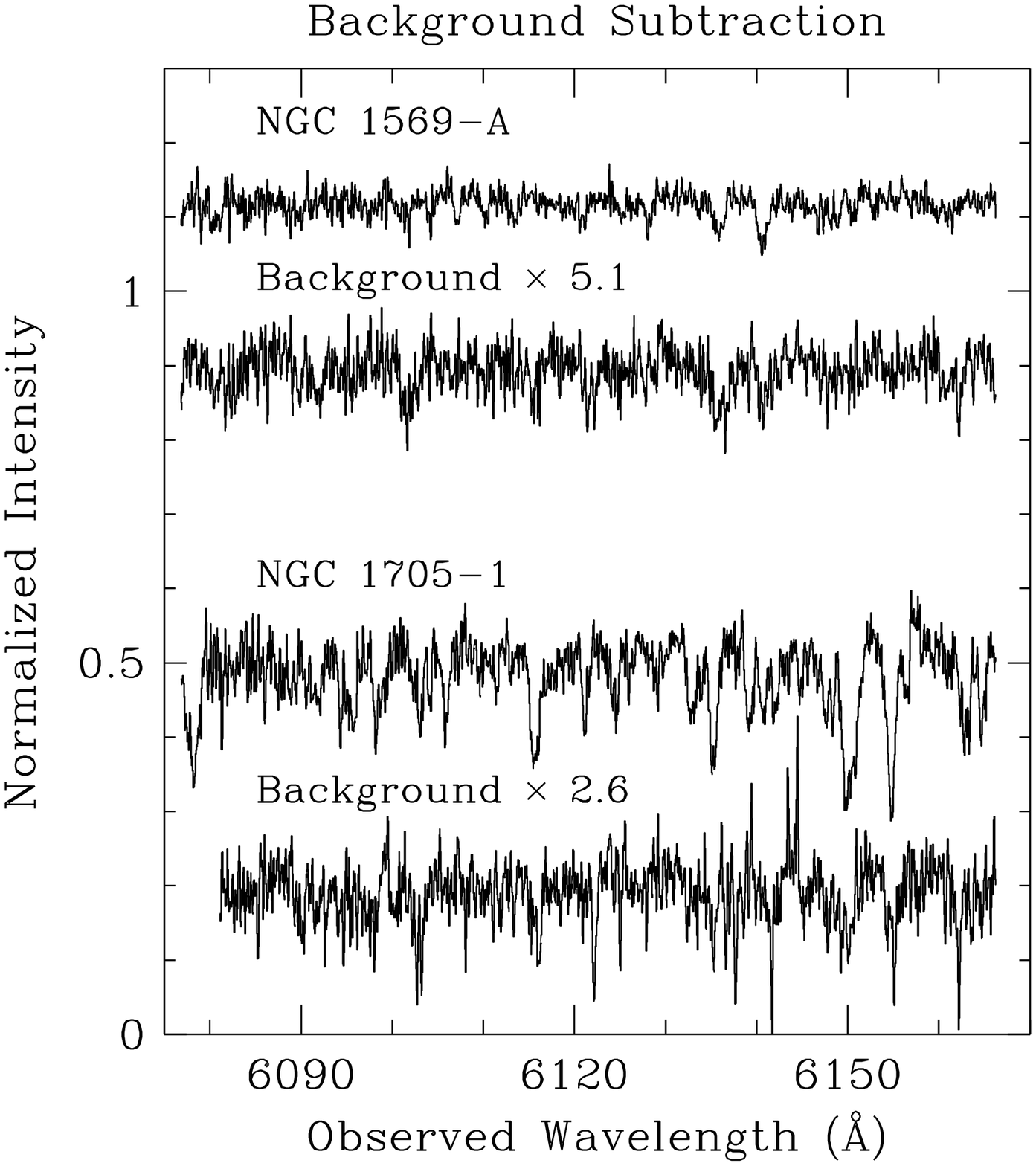}
\caption{}
\end{figure}

\clearpage
\begin{figure}
\figurenum{5{\it a}}
\plotone{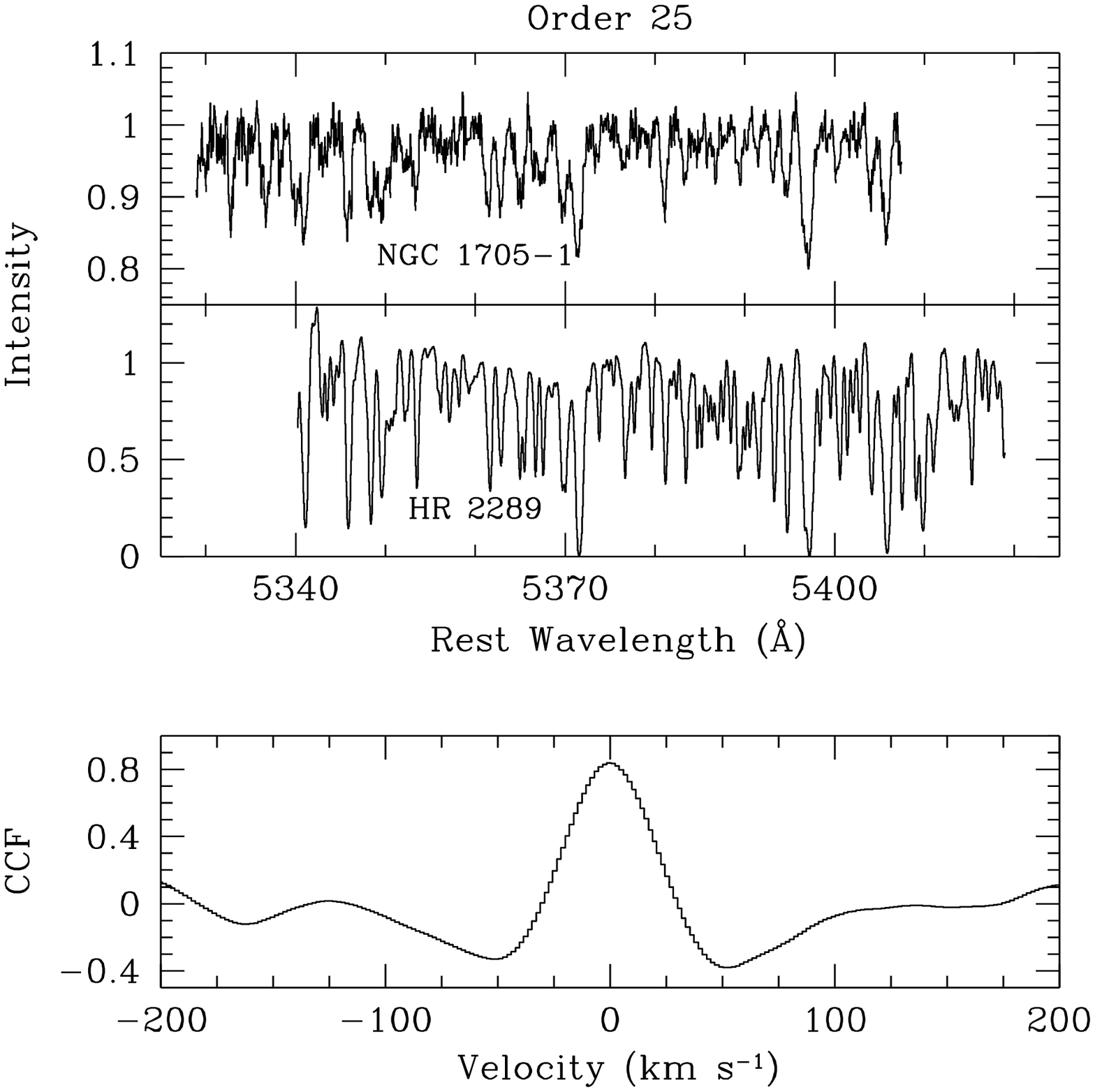}
\caption{}
\end{figure}

\clearpage
\begin{figure}
\figurenum{5{\it b}}
\plotone{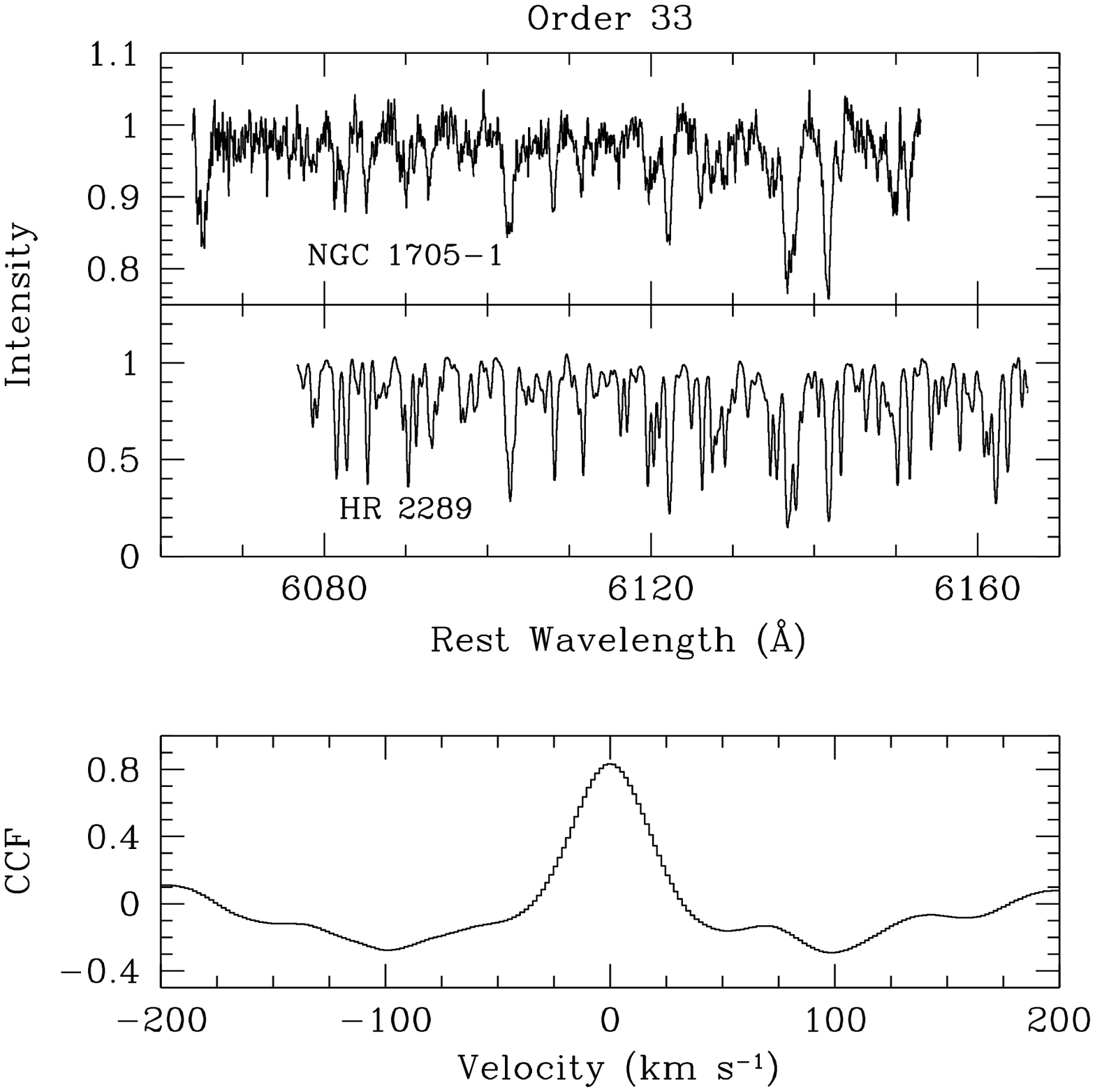}
\caption{}
\end{figure}

\clearpage
\begin{figure}
\figurenum{6{\it a}}
\plotone{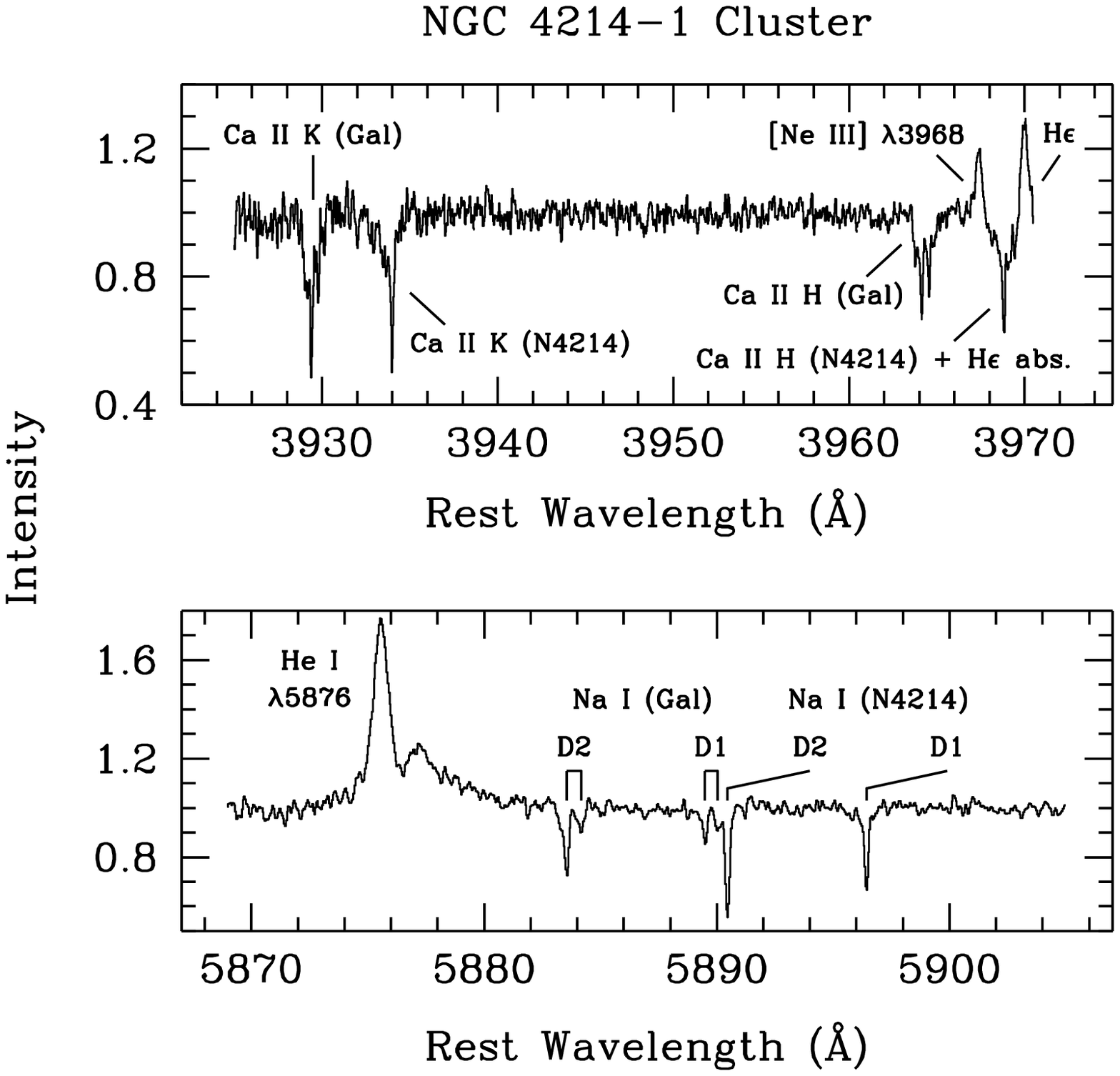}
\caption{}
\end{figure}

\clearpage
\begin{figure}
\figurenum{6{\it b}}
\plotone{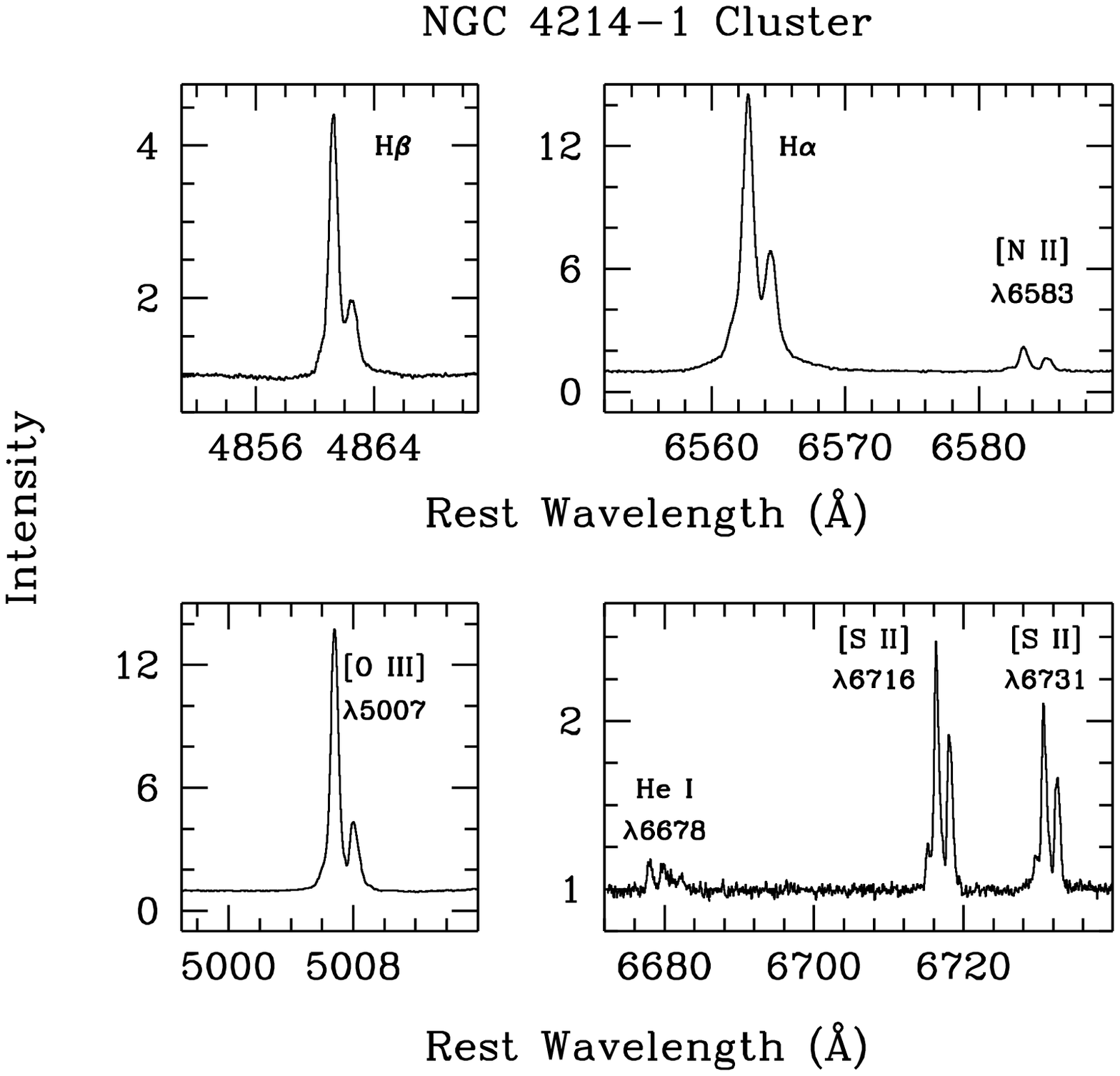}
\caption{}
\end{figure}

\end{document}